\documentclass{aastex631}

\usepackage{pifont}
\usepackage{amsmath}
\usepackage{multirow}
\usepackage{bm}
\usepackage{multirow}
\usepackage{tabularx}
\usepackage{longtable,tabu}

\newcommand{\tm}[1]{{\bf \textcolor[rgb]{0.59, 0.29, 0.0}{#1}}}
\newcommand{\be}{\begin{equation}}
\newcommand{\ee}{\end{equation}}
\newcommand{\bea}{\begin{eqnarray}}
\newcommand{\eea}{\end{eqnarray}}

\AtBeginDocument{
  \let\namerefOld\nameref
  \renewcommand{\nameref}[1]{\textit{\namerefOld{#1}}}
}

\usepackage{xcolor}

\definecolor{color1}{HTML}{440154}
\definecolor{color2}{HTML}{481568}
\definecolor{color3}{HTML}{482677}

\definecolor{color18}{HTML}{B8DE29}
\definecolor{color19}{HTML}{DCE318}
\definecolor{color20}{HTML}{FDE725}

\shorttitle{}
\shortauthors{Debanjan et al.}
\graphicspath{{./}{figure/}}

\begin{document}


\title{Analysis of Neutron Star $f-$mode Oscillations in General Relativity with Spectral Representation of Nuclear Equations of State}

\correspondingauthor{Sarmistha Banik}

\author[0009-0004-5222-6690]{Debanjan Guha Roy}
\affiliation{Department of Physics, Birla Institute of Technology and Science Pilani, Hyderabad Campus, Telangana 500078, India.}

\author[0000-0003-2633-5821]{Tuhin Malik}
\affiliation{CFisUC, Department of Physics, University of Coimbra, 3004-516 Coimbra, Portugal.}

\author[0009-0008-0415-373X]{Swastik Bhattacharya}
\affiliation{Department of Physics, Birla Institute of Technology and Science Pilani, Hyderabad Campus, Telangana 500078, India.}

\author[0000-0003-0221-3651]{Sarmistha Banik}
\affiliation{Department of Physics, Birla Institute of Technology and Science Pilani, Hyderabad Campus, Telangana 500078, India.}
\email{sarmistha.banik@hyderabad.bits-pilani.ac.in}

\begin{abstract}
We study quasinormal $f-$mode oscillations in neutron star(NS) interiors within the linearized General Relativistic formalism. We utilize approximately 9000 nuclear Equations of State (EOS) using spectral representation techniques, incorporating constraints on nuclear saturation properties, chiral Effective Field Theory ($\chi$EFT) for pure neutron matter, and perturbative Quantum
Chromodynamics (pQCD) for densities pertinent to NS cores. The median values of f-mode frequency, $\nu_f$ (damping time, $\tau_f$) for NS with masses ranging
from 1.4 - 2.0 $M_\odot$ lie between 1.80 - 2.20 kHz (0.13 - 0.22 s) for our
entire EOS set. Our study reveals a weak correlation between $f-$mode
frequencies and individual nuclear saturation properties, prompting the
necessity for more intricate methodologies to unveil multi-parameter
relationships. We observe a robust linear relationship between the radii and $f-$mode frequencies for different NS masses. Leveraging this correlation alongside NICER observations of PSR J0740+6620 and PSR J0030+0451, we establish constraints that exhibit partial and minimal overlap for observational data from Riley et al. and Miller et al. respectively with our nucleonic EOS dataset. Moreover, NICER data aligns closely with radius and frequency values for a few hadron-quark hybrid EOS models. This indicates the need to consider additional exotic particles such as deconfined quarks at suprasaturation densities. We conclude that future observations of the radius or $f-$mode frequency for more than one NS mass, particularly at the extremes of viable NS mass scale, would either rule out nucleon-only EOS or provide definitive evidence in its favour.
\end{abstract}


\section{Introduction}
\label{sec:intro}
\noindent

Neutron stars (NS) are among the densest objects existing in the present universe with a measured and estimated mass in the range of 1 - 2$M_\odot$, confined in a sphere of radius $\sim$ 10 km. These tiny but massive stars harbor matter  
under extreme physical conditions, 
that remains inaccessible to direct observation. Hence, our knowledge of a NS's interior, including the behavior of matter at such extreme densities, remains theoretical and is the subject of ongoing research and modeling. However, the new generations of telescopes have provided valuable observation data in the recent past and helped refine the theoretical description - called the equation of state (EOS). The EOS of matter inside a NS plays a critical role in determining various properties of NS, such as its mass, radius, moment of inertia, tidal deformability, oscillation frequencies and the structure of the internal layers. It is essential for understanding the observational characteristics of NS, including their gravitational wave signals, electromagnetic radiation, and thermal emissions.  The maximum observed mass of NS  $\sim 2M_{\odot}$  ruled out the softer EOS \citep{Cromartie2019, Fonseca2021, Demorest2010, Antoniadis2013}.
The International Space Station (ISS) based X-ray telescope and instrument package, NS Interior Composition Explorer (NICER) \citep{Gendreau2017} measures the mass and radius of a pulsar simultaneously for the first time. Electromagnetic radiation from the hotspots at the magnetic poles of a rotating NS provides us with a direct route of observation when the radiation crosses our line of sight \citep{Miller2021, Riley2021}. 
The detection of gravitational waves from NS mergers events have provided stringent limits on tidal deformability for a NS with canonical mass of $1.4M_{\odot}$, adding valuable insights into the EOS \citep{Abbott1_2018}.
The nuclear physics data at the moderate densities combined with recent NICER observations can further improve the constraints.
Additionally, new X-ray observatories like the European Space Agency's Athena mission \citep{Nandra2013} and NASA's Lynx mission \citep{Kouveliotou2014} are expected to offer even higher-resolution X-ray observations of NS. The third generation Einstein Telescope (ET) \citep{Branchesi_Einstein2023} will have a sensitivity ten times better compared to the second generation detectors.  Advanced LIGO  and  Virgo would be optimized for the detection of gravitational signals at low frequency. Ongoing and future surveys, such as the Square Kilometre Array (SKA) \citep{Kramer2015, Watts2015}, is poised to discover more pulsars and provide a better understanding of the population of NS, their properties. The coordinated observations of NS mergers in multiple wavelengths, such as X-rays, gamma-rays, and radio waves will definitely help enhance our understanding of the EOS of dense matter.

The NS core is believed to be made up of neutrons, along with an admixture of protons and leptons. There are several EOS models, that can be broadly categorised into non-relativistic and relativistic models. Among them, the relativistic mean field models (RMF) varieties have been successfully  applied to study NS properties as well as  finite nuclei properties. Over the decades additional mesons and  non-linear self-interaction as well as cross-coupling terms for all the mesons have been added to the original RMF i.e. \cite{Walecka1974} Model. Various sets of tabulated RMF EOS, compatible with recent astrophysical observations \citep{Nandi2019} are available\footnote{CompOSE website (https://compose.obspm.fr)} depending upon the form of interactions in the Lagrangian density. The associated couplings for the interactions are fitted to various nuclear matter parameters such as  binding energy per particle, incompressibility, the symmetry energy coefficients and their derivatives, that are determined from terrestrial experiments at nuclear saturation density. Nevertheless the EOS provides a reasonable description of the nuclear matter at suprasaturation densities. The nuclear matter parameters define the NS EOS and are correlated to NS observables \citep{Agrawal2021, Ghosh2022, Malik2023, Pradhan2023_nuc}.
At this high density, different models  consider the possibility of strange matter like hyperons, quarks, Bose-Einstein condensates, under extreme pressures and densities also \citep{Banik2014, Char2015, Malik2021}. It's a challenge to unravel the NS core composition and precisely determine its EOS owing to our limited knowledge of matter under extreme physical conditions.

In the absence of any universally-acceptable model EOS, subject to limited observational data, several recent works have used the Spectral representations of realistic EOS as an accurate and efficient way of parametrizing the high-density sections of the NS EOS. \citep{Abbott1_2018, Miller2019, Raaijmakers2020}. This is a powerful theoretical tool, to describe the EOS in terms of the spectral functions, incorporating our current knowledge of fundamental physics. 
The tabulated version of the realistic EOS 
often infuses numerical error in the calculation of the relevant thermodynamic quantities 
\citep{Servignat_arxiv2023}. 
A robust analytic representation of the EOS like the Spectral representation \citep{Lindblom2010, Lindblom2018, Lindblom2022}, the generalized piecewise polytropic representation \citep{OBoyle2020} or the speed of sound parametrization \citep{Tews2018, Greif2019} becomes quite useful in scenarios where we need numerical determination of thermodynamic quantities  associated with a cold $\beta-$equilibrated EOS at any arbitrary energy density inside the NS.

The structure parameters like mass, radius, tidal deformability have taken crucial roles to address the uncertainties in NS EOS. The quasi-periodic oscillations (QPOs) are also valuable tools for studying the compact objects and the processes occurring in their binary systems. NS can exhibit a variety of oscillation modes mainly observed in the X-ray light curves of NS in various types of binary systems, including low-mass X-ray binaries (LMXBs) and NS X-ray binaries.  A General Relativistic description of oscillations due to non-radial perturbations involves what are called the quasinormal modes (QNM). Einstein equations govern the dynamics both inside and outside the star. QNMs are different than the normal modes describing Newtonian oscillations of matter inside the star. These modes are damped due to the emission of gravitational waves (GW) from the star. The emitted GWs, given by the perturbations of the spacetime metric tensor, carry information about the internal structure and composition of the NS. The real part of the complex-valued QNM frequency represents the actual frequency of the oscillations, and the imaginary part corresponds to an exponential damping of the wave. Outside the star, the determination of frequency and damping time of the QNMs become an eigenfrequency problem subject to appropriate boundary conditions both at the surface of the NS and at spatial infinity. The internal structure and dynamics of the matter inside it influence the frequencies and damping times of the QNMs via the boundary condition at the surface of the NS. Inside the star, the oscillations of the matter and the GWs are coupled to each other via the Einstein equations. 

The study of QNMs of oscillating astrophysical objects started with spherically symmetric black hole spacetimes [See \citep{PhysRev.108.1063, 1970Natur.227..936V, PhysRevLett.24.737}] and was later extended to compact astrophysical stars\citep{Thorne1967} including NS
[See the review \citep{1999LRR.....2....2K} for references]. The modes of oscillation inside the NS are characterized based on the physics behind the restoring force. For example, the pressure of the fluid (which is generally considered incompressible) in the NS interior, acts as the restoring force for $p-$modes. Buoyancy restores the so-called gravity $g-$modes which originates from internal temperature or composition gradients. The most fundamental mode of oscillation is the $f-$modes which arise out of density perturbations. They typically have a frequency of around 2.4 kHz \citep{Rezzolla2003_review, kokkotas2001inverse} and have the strongest coupling with the emission of GWs.
Determination of the eigen-frequencies while ignoring the metric perturbations during fluid oscillations (relativistic Cowling approximation) to simplify the calculations overestimates the frequency of certain fundamental modes [See the discussion on the accuracy of this approximation in \citep{PhysRevD.102.063025}]. We can improve the accuracy by introducing the emission of GWs and using a post-Newtonian approximation to estimate the damping rate of the f-modes as a result. This way, one can obtain scaling relations between bulk properties of the star, e.g., compactness, mass, moment of inertia and mode frequency, and the damping time \citep{Andersson2019, andersson1996, andersson1998, Benhar2004, detweiler1975variational, kokkotas2001inverse}. When the problem is treated General Relativistically, we get $f-$modes for the QNM.  

Now, several works in the literature have demonstrated that for $f-$modes, the relativistic analysis also leads to very similar scaling relations. Such scaling relations can thus be taken as universal[See for example \citep{andersson1996, andersson1998, lattimer2001neutron, yagi2013love, tsui2005universality, lau2010inferring}]. This feature of universality can be used as a landmark. One can check how well the calculated frequencies and damping times of the QNMs for a given EOS fit the scaling relations as a sort of consistency requirement[For details, see \citep{Andersson2021}]. This is a strategy we consistently follow throughout this paper. Another feature of the NS, the tidal deformability, also encodes important information about the shape of the star. 
An external tidal gravitational field acting on the NS contributes to the quadrupole and higher moments of the matter distribution inside[For more details, see \citep{andersson2020exploring, pratten2020gravitational, andersson2021phenomenology, poisson2014gravity, chan2014multipolar}].

Apart from the $f-$modes, there are also other modes that typically correspond to one or more additional physical processes. These modes mostly have frequencies higher than that of $f-$modes. For example, the frequency of the g-modes, which appear when the temperature and/or the pressure of the NS vary from one point inside the star to another, can lie in a range starting from the order of 100 kHz \citep{Reisenegger1992}. For this reason, one expects $g-$modes to appear when phase transition occurs inside the NS \citep{Andersson2019}. Apart from these, $p-$modes and $r-$modes can also occur due to pressure and spin-related instabilities in the NS, respectively [For details, see the review \citep{Andersson2019} and the references therein]. However, in this work, we focus only on $f-$modes for the NS, determining their frequencies and damping times from fully general relativistic calculations.  

Recent publications \citep{Sotani2021, Pradhan2022, Kunji2022, Pradhan2023} show efforts to estimate nuclear matter parameters using stellar oscillation modes. \citep{Pradhan2022} reported how the effective nucleon mass ($m^\star$) at saturation is correlated with the $f-$mode frequencies for a set of EOS generated within the relativistic mean field formalism. A strong correlation of the stellar oscillation frequencies with the symmetry energy parameters like $J_{sym, 0}$, $K_{sym, 0}$, $L_{sym, 0}$, and so on would allow us to infer the composition of matter at very high densities in the NS interior \citep{Patra2023}.

In this work we have prepared a set of $\sim$9000 nucleonic EOS though Bayesian analysis of relativistic mean field model parameters. Constraints of recent interest like the $\chi$EFT and pQCD at relevant energy densities are imposed to infer the parameter set. Then we determine a highly accurate Spectral representation of the generated set and numerically solve the fully General Relativistic perturbation equations for a set of equilibrium configuration NS with varying mass for each EOS. We report the corresponding $f-$mode frequencies and the gravitational wave damping times over the EOS set and finally explore correlation of the frequencies and damping times with nuclear matter parameters and stellar observables.

 The paper is organised in the following manner. Sec. \ref{sec:formalism} provides a brief overview of the relativistic mean field formalism for the nuclear EOS at zero temperature followed by a brief description framework for spectral fitting of the EOS, and the formalism for non-radial quadrupolar oscillations of a non-rotating NS. Sec. \ref{sec:results} contains the results for the set of parametrized spectral version of the entire EOS set. We discuss the results and conclude our work in Sec. \ref{sec:conclusion}.

\begin{figure*}
    \includegraphics[width=\textwidth]{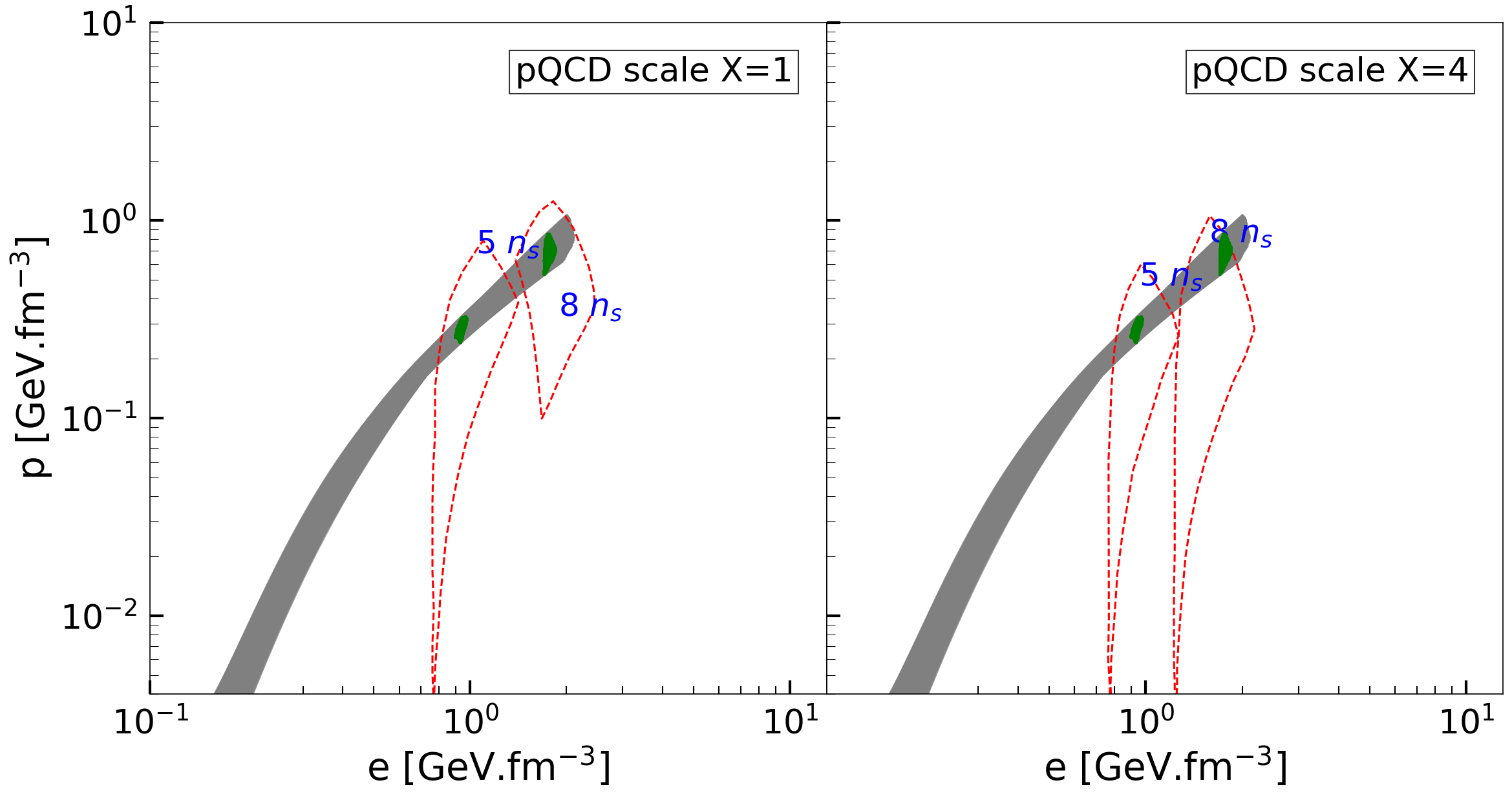}
    \caption{We display the  pressure and energy density for the employed EOS set, enclosing areas compliant with pQCD-derived restrictions for baryon number densities equal to 5$n_s$ and 8$n_s$ (where $n_s$=0.16 fm$^{-3}$) \protect\citep{Komoltsev2022} by red dashed lines, informed by the stringent renormalization scale parameter, $X=1$ ($X=4$), according to \protect\citep{Kurkela2010} in the left panel (right panel). We indicate the energy densities and pressures that satisfy the pQCD constraints by green patches for the corresponding number densities. It will be significant to note that the central density for the maximum mass NS within 7$n_s$  across our entire EOS set.}
    \label{fig:EOS_pQCD}
\end{figure*}

\section{Formalism}
\label{sec:formalism}
\subsection{RMF EOS} \label{RMF}
 We generate a set of EOS 
 within a  description of nuclear matter based on a relativistic field theoretical approach 
 where the nuclear interaction between nucleons is introduced through the exchange of the scalar-isoscalar meson $\sigma$, the vector-isoscalar meson $\omega$, and the vector-isovector meson $\varrho$. 
 
The Lagrangian density is given by \citep{Dutra2014, Malik2023}
        \begin{equation}
          \mathcal{L}=   \mathcal{L}_N+ \mathcal{L}_M + \mathcal{L}_{NL} +\mathcal{L}_{leptons}
        \end{equation} 
where $$\mathcal{L}_{N} = \bar{\Psi}\Big[\gamma^{\mu}\left(i \partial_{\mu}-g_{\omega} \omega_\mu - \frac{1}{2}g_{\varrho} {\boldsymbol{t}} \cdot \boldsymbol{\varrho}_{\mu}\right) - \left(m_N - g_{\sigma} \sigma\right)\Big] \Psi$$ denotes the Dirac equation for the nucleon doublet (neutron and proton) with bare mass $m_N$, $\Psi$ is a Dirac spinor, $\gamma^\mu $ are the Dirac matrices, and $\boldsymbol{t}$ is the isospin operator. $\mathcal{L}_{M}$ represents the mesons, given by
\begin{eqnarray}
\mathcal{L}_{M}  &=& \frac{1}{2}\left[\partial_{\mu} \sigma \partial^{\mu} \sigma-m_{\sigma}^{2} \sigma^{2} \right] - \frac{1}{4} F_{\mu \nu}^{(\omega)} F^{(\omega) \mu \nu} + \frac{1}{2}m_{\omega}^{2} \omega_{\mu} \omega^{\mu}   \nonumber \\
  &-& \frac{1}{4} \boldsymbol{F}_{\mu \nu}^{(\varrho)} \cdot \boldsymbol{F}^{(\varrho) \mu \nu} + \frac{1}{2} m_{\varrho}^{2} \boldsymbol{\varrho}_{\mu} \cdot \boldsymbol{\varrho}^{\mu} \nonumber
\end{eqnarray}
where $F^{(\omega, \varrho)\mu \nu} = \partial^ \mu A^{(\omega, \varrho)\nu} -\partial^ \nu A^{(\omega, \varrho) \mu}$ are the vector meson  tensors.
\begin{eqnarray}
\mathcal{L}_{NL}&=&-\frac{1}{3} b~m_N~ g_\sigma^3 (\sigma)^{3}-\frac{1}{4} c (g_\sigma \sigma)^{4}+\frac{\xi}{4!} g_{\omega}^4 (\omega_{\mu}\omega^{\mu})^{2}  \nonumber \\
&+&\Lambda_{\omega}g_{\varrho}^{2}\boldsymbol{\varrho}_{\mu} \cdot \boldsymbol{\varrho}^{\mu} g_{\omega}^{2}\omega_{\mu}\omega^{\mu} \nonumber
\end{eqnarray}
contains the non-linear terms with parameters $b$, $c$, $\xi$, $\Lambda_{\omega}$ to take care of the high-density behavior of the matter. 
$g_i$'s are the couplings of the nucleons to the meson fields $i = \sigma, \omega, \varrho$, with masses $m_i$.
\noindent
Finally, the Lagrangian density for the  leptons is given as $\mathcal{L}_{leptons}= \bar{\Psi_l}\Big[\gamma^{\mu}\left(i \partial_{\mu}  
-m_l \right)\Psi_l\Big]$, where $\Psi_l~(l= e^-, \mu^-)$ denotes the lepton spinor; leptons are considered non-interacting.
\noindent
The equations of motion for the mesons  fields are obtained from Euler-Lagrange equations:
		\begin{eqnarray}
			{\sigma}&=& \frac{g_{\sigma}}{m_{\sigma,{\rm eff}}^{2}}\sum_{i} \rho^s_i\label{sigma}\\
			{\omega} &=&\frac{g_{\omega}}{m_{\omega,{\rm eff}}^{2}} \sum_{i} \rho_i \label{omega}\\
			{\varrho} &=&\frac{g_{\varrho}}{m_{\varrho,{\rm eff}}^{2}}\sum_{i} t_{3} \rho_i, 
		\end{eqnarray}
 where $\rho^s_i$ and $\rho_i$ are, respectively, the scalar density and the number density of nucleon $i$. The effect of the nonlinear terms on the magnitude of the meson fields is clearly shown through the following terms
 \begin{eqnarray}
   m_{\sigma,{\rm eff}}^{2}&=& m_{\sigma}^{2}+{ b g_\sigma^3}{\sigma}+{c g_\sigma^4}{\sigma}^{2} \label{ms} \\ 
    m_{\omega,{\rm eff}}^{2}&=& m_{\omega}^{2}+ \frac{\xi}{3!}g_{\omega}^{4}{\omega}^{2} +2\Lambda_{\omega}g_{\varrho}^{2}g_{\omega}^{2}{\varrho}^{2}\label{mw}\\
    m_{\varrho,{\rm eff}}^{2}&=&m_{\varrho}^{2}+2\Lambda_{\omega}g_{\omega}^{2}g_{\varrho}^{2}{\omega}^{2}, \label{mr}
 \end{eqnarray}

 and are solved using the relativistic mean-field approximation with the constraint of $\beta$-equilibrium i.e. $\mu_n-\mu_p=\mu_e$, where $\mu$'s are chemical potential of the respective particles. Also, the total baryon number ($n_p+n_n$)  conservation and the charge neutrality condition $n_p =n_e$ are enforced.

The pressure and energy density of the baryons and 
leptons are given by the following expressions:
\begin{equation}
\begin{aligned}
\epsilon &= \sum_{i=n,p,e,\mu}\frac{1}{\pi^2}\int_0^{k_{Fi}} \sqrt{k^2+m_i^*}\, k^2\, dk \\
&+ \frac{1}{2}m_{\sigma}^{2}{\sigma}^{2}+\frac{1}{2}m_{\omega}^{2}{\omega}^{2}+\frac{1}{2}m_{\varrho}^{2}{\varrho}^{2}\\
&+ \frac{b}{3}(g_{\sigma}{\sigma})^{3}+\frac{c}{4}(g_{\sigma}{\sigma})^{4}+\frac{\xi}{8}(g_{\omega}{\omega})^{4} + \Lambda_{\omega}(g_{\varrho}g_{\omega}{\varrho}{\omega})^{2},
\end{aligned}
\end{equation}
   where $m_i^*=m_i-g_{\sigma} \sigma$ for protons and neutrons and $m_i^*=m_i$ for electrons and muons, and $k_{Fi}$ is the Fermi moment of particle $i$. 
The pressure is determined from the thermodynamic relation for every component:
\begin{equation}
p = \sum_{i}\mu_{i}\rho_{i}-\epsilon.
\end{equation}

We generate our EOS set using Bayesian setup with minimal constraints as done in \cite{Malik2023}. These constraints include nuclear saturation properties such as a saturation density of $\rho_0=0.153 \pm 0.005$ fm$^{-3}$, a binding energy per nucleon of $e_0=-16.1 \pm 0.2$ MeV, nuclear incompressibility of $K_0 = 230 \pm 40$ MeV, and a symmetry energy of $J_{sym,0} = 32.5 \pm 1.8$ MeV at the saturation density. Furthermore, we impose a low density constraint for the pure neutron matter pressure from chiral effective field theory ($\chi$EFT) \citep{Hebeler2010} with twice the uncertainty and demand that the maximum NS mass corresponding to the EOS is above the observed $2M_{\odot}$. Additionally, we employ the perturbative Quantum Chromodynamics (pQCD) derived constraint on the pressure at eight times the baryon density ($n_s=0.16$ fm$^{-3}$) for a pQCD scale $X=4$ \citep{Komoltsev2022}. Finally, we choose $\sim$9,000 nucleonic EOS for the present calculation from the posterior. 

\subsection{Spectral representation of EOS}
\label{subsec:spectral}

\begin{figure*}
\includegraphics[width=\textwidth]{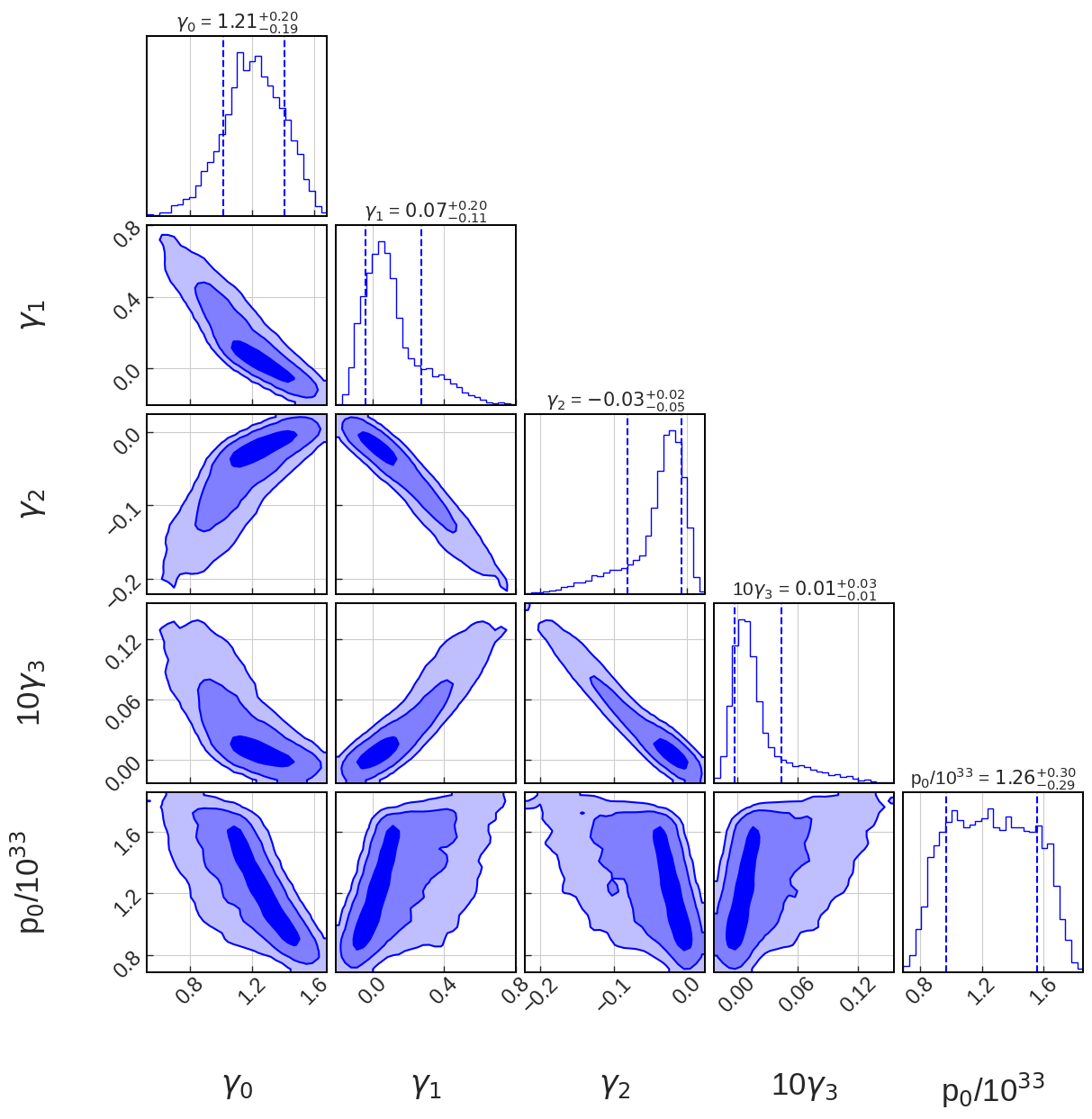}
\caption{The corner plot shows the distributions of parameters $\gamma_0$, $\gamma_1$, $\gamma_2$, $\gamma_3$, and $p_0$ \tm{for} the spectral representation of the EOS within the acquired set of RMF EOS for this study. Vertical lines indicate the 68\% credible intervals (CIs), while varying shades in the 2D distribution denote the 1$\sigma$, 2$\sigma$, and 3$\sigma$ CIs, respectively.}
\label{fig:corner_spectral}
\end{figure*}

We find the spectral parameters for our set of EOS by fitting the tabular RMF EOS with a spectral version.
The methodology in \cite{Lindblom2010} leads us to faithful representations of realistic EOS for nuclear matter inside the NS by constructing spectral expansions of key thermodynamic quantities.
Linear combinations of a complete set of functions like the Fourier basis functions allow us to expand a thermodynamic quantity like the adiabatic index, defined as
\begin{equation}
    \Gamma(p) = \dfrac{\epsilon+ p}{p} \dfrac{dp}{d \epsilon}.
\end{equation}
The coefficients of such basis functions in the expansion uniquely determine any given physically valid EOS. One can find that such a representation satisfies the basic necessary thermodynamic conditions.
\noindent
To obtain a pressure-based form, we expand the adiabatic index $\Gamma(p)$ in terms of the basis functions $\Phi_k (p)$ as
\begin{equation}
  \Gamma(p)=\exp\left[\sum_k\gamma_k\Phi_k(p)\right].
\end{equation}
We choose $\Phi_k(p) = \Phi_k(x)$, where $x$ is a dimensionless logarithmic pressure variable $x = log(p/p_0)$ as in \cite{Lindblom2010}, $p_0$ being the minimum pressure where the high-density part of the EOS is joined to the low-density part of BPS\citep{Baym1971} EOS. The $\gamma_k$'s are the spectral coefficients.
\noindent
Numerical integration of the following gives us $\epsilon(p)$ for the obtained expansion of $\Gamma(p)$
\begin{equation}
  \epsilon(p)=\frac{\epsilon_0}{\mu(p)}+\frac{1}{\mu(p)}\int_{p_0}^p
  \frac{\mu(p')}{\Gamma(p')}dp',
\end{equation}
where $\mu(p)$ is the chemical potential defined as
\begin{equation}
  \mu(p)=\exp\left[-\int_{p_0}^p \frac{dp'}{p'\Gamma(p')}\right].
\end{equation}
The energy density $\epsilon_0$, at the minimum pressure $p_0$ is the constant of integration. We match low-density EOS at this pressure generally occurring at a number density slightly below the nuclear saturation density such that there is no unexpected phase transition at the junction.
One can determine an empirical fit to any given EOS with arbitrary accuracy by increasing the number of basis functions used. There are 4 spectral coefficients corresponding to four basis functions as in \cite{Lindblom2010} for fitting the set of RMF EOS we are using. We plot distribution of the spectral parameters $\gamma_0$, $\gamma_1$, $\gamma_2$, $\gamma_3$, $p_0$ in Fig. \ref{fig:corner_spectral}. Apart from the coefficients, parameters $\epsilon_0$ and $x_{max}$ have fixed values.
We optimize the choice of spectral coefficients, $\gamma_k$, by minimizing the differences between $\epsilon_{fit}(x_i, \gamma_k)$ and the energy density $\epsilon_i = \epsilon(x_i)$ values from the tabular version of EOS for a specific set of $x$'s (or $p$'s). The following residue quantifies the faithfulness of the fitting
\begin{equation}
    \Delta^2(\gamma_k) = \sum ^N _{i=1} \frac{1}{N} \biggl\{ log \biggl[ \dfrac{\epsilon_{fit}(x_i, \gamma_k)}{\epsilon_i} \biggr] \biggr\}^2
    \label{e:residue_spectral}
\end{equation}
where we sum over the chosen pressure values between $p_0 \leq p_i \leq p_{max}$; $p_{max}$ is the pressure at the center of the non-rotating maximum mass NS for the given EOS. The spectral version of the EOS serves as the input for the Tolman-Oppenheimer-Volkoff equations, which gives us the equilibrium configuration of the star.
\subsection{Metric with perturbations}
\label{subsec:metric_perturbations}

To determine the frequencies of the f-modes, we solve the Einstein equations $G_{\mu\nu}= 8\pi G T_{\mu\nu}$. 
We assume here that the gravitational waves are perturbations of the static background spacetime metric for the non-rotating NS. \noindent
The perturbed metric has the form:
\begin{equation}
  g_{\mu\nu} = g_{\mu\nu}^0 + h_{\mu\nu},
  \label{scw_pert}
\end{equation}
The time dependence of the perturbed metric components can be expressed by the factor $e^{i\omega t}$ for a wave mode. Here $\omega$ is complex as the waves will decay due to the imposed open boundary conditions, which will be discussed shortly. The real part of $\omega$ is the oscillation frequency, while the imaginary part gives the inverse of the gravitational wave damping time (if it is positive) of the wave mode. These oscillations are studied considering the linearized Einstein equations, coupled with the equations of hydrodynamics, for our set of EOS, with suitably posed boundary conditions. A small perturbation $h_{\mu\nu}$ is assumed on a static spherically symmetric background metric given by
\begin{align}
ds^2 = &-e^{\nu(r)} \left[ 1 + r^l H_0(r) e^{i \omega t} Y_{lm}(\phi, \theta) \right] c^2 dt^2 \nonumber \\
+&e^{\lambda(r)} \left[ 1 - r^l H_0(r) e^{i \omega t} Y_{lm}(\phi, \theta) \right] dr^2 \nonumber \\
+& \left[ 1 - r^l K(r) e^{i \omega t} Y_{lm}(\phi, \theta) \right] r^2 d\omega^2 \nonumber \\
-& 2 i \omega r^{l+1} H_1(r) e^{i \omega t} Y_{lm}(\phi, \theta) dt dr,
\end{align}

Here $e^{\lambda(r)} = \frac{1}{1 - \frac{2Gm(r)}{c^2r}}$ and $e^{\nu(r)} = exp \left( \frac{2G}{c^2} \int_{0}^{r} \left\{\dfrac{m(r') + \frac{4 \pi p(r')r'^3}{c^2}}{r' \left[ r' - \frac{2m(r')G}{c^2} \right] } \right\} dr' \right) e^{\nu_0}$; $m(r')$ and $p(r')$ being the enclosed mass and pressure of the star at $r'$. Here $H_0$, $H_1$, and $K$  are radial perturbations of the metric and the angular part is contained in the spherical harmonics $Y_{m}^l$ with $l$ denoting the orbital angular momentum number and $m$ the azimuthal number.
\subsection{Lindblom-Detweiler equations}
The perturbations of the energy-momentum tensor of the fluid also need to be taken into account in the Einstein equation. The components of Lagrangian displacement vector $\xi^a(r, \theta, \phi)$ describes the perturbations of the fluid inside the star:
\begin{align}
    \xi^r &= r^{l-1} e^{-\lambda/2} W Y^l_m e^{i \omega t} \\
    \xi^{\theta} &= -r^{l-2} V \partial_{\theta} Y^l_m e^{i \omega t} \\
    \xi^{\phi} &= - \dfrac{r^{l-2}}{sin^2 \theta} V \partial_{\theta} Y^l_m e^{i \omega t},
\end{align}
where $W$ and $V$ are functions with respect to r that describe fluid perturbations. Fluid perturbations exist only inside the star.
The introduction of a new variable $X$ leads to simplification of perturbation equations for NS interior. $X$ is related to the rest of the perturbation functions through the following equations
\begin{eqnarray}
    H_0 &=& \Bigl\{ 8 \pi r^2 e^{-\nu/2} X - \left[ (n+1)Q \right] \nonumber \\
    &-& \omega^2 r^2 e^{-(\nu + \lambda)} H_1 + \left[ n - \omega^2 r^2 e^{-\nu} \right. \nonumber \\
    &-& \left. e^{\lambda}Q (Q - e^{-\lambda}) \right] K \Bigr\} (2b + n + Q)^{-1}, \\
    V &=& \left[ \dfrac{X}{\epsilon + p} - \dfrac{Q}{r^2} e^{(\nu + \lambda)/2} W - e^{\nu /2} \dfrac{H_0}{2} \right] \dfrac{e^{\nu /2}}{\omega^2},
\end{eqnarray}
where $n = (l-1)(l+2)/2, b = Gm/rc^2, Q = b + 4 \pi Gr^2p/c^4,$ and $\epsilon$ is the local energy density.
The gravitational wave equations can be written as a set of four coupled linear differential equations for the four perturbation functions: $H_1$, $K$, $W$, and $X$, which do not diverge inside the star for any given value of $\omega$ \citep{Lindblom1983}.

\begin{align}
    r \frac{dH_1}{dr} &= - \left[ l + 1 + 2be^{\lambda} 4 \pi r^2 e^{\lambda} (p - \epsilon) \right] H_1 \nonumber \\
    & \quad + e^{\lambda} \left[ H_0 + K - 16 \pi (\epsilon + p) V \right], \\
    r \frac{dK}{dr} &= H_0 + (n+1)H_1 + \left[ e^{\lambda}Q - l - 1 \right] K \nonumber \\
    & \quad - 8 \pi (\epsilon + p) e^{(\lambda/2)} W, \\
    r \frac{dW}{dr} &= -(l+1) \left[ W + le^{\lambda/2}V \right] \nonumber \\
    & \quad + r^2e^{\lambda/2} \left[ \frac{e^{-\nu/2}X}{(\epsilon+p)c^2_{eq}} + \frac{H_0}{2} + K \right], \nonumber \\
    r \frac{dX}{dr} &= -lX + \frac{(\epsilon + p) e^{(\nu/2)}}{2} \biggl\{ (3e^{\lambda}Q - 1)K \nonumber \\
    & \quad - \frac{4(n+1)e^{\lambda}Q}{r^2} V + (1-e^{\lambda}Q)H_0 \nonumber \\
    & \quad + (r^2 \omega^2 e^{-\nu} + n + 1)H_1 \nonumber \\
        & \quad - \left[ 8 \pi ( \epsilon+p)e^{ \lambda/2} + 2\omega^2 e^{\lambda/2 - \nu} r^2 \right. \nonumber \\
    & \quad \left. \times \frac{d}{dr} \left( \frac{e^{-\lambda/2}}{r^2} \frac{d\nu}{dr} \right) \right] W \biggr\}.
\end{align}

where $c^2_{eq} = dp/d\epsilon$ is the equilibrium speed of sound of NS matter undergoing oscillations.
\subsection{Boundary conditions for perturbation equations}
We discuss the boundary conditions imposed here to solve the Einstein equations inside the star. For that, one needs to integrate the perturbation equations from the center of the star to a certain radius inside the star. The boundary condition for the perturbation functions at the center of the star $r=0$ are

\begin{align}
    \label{LD_bc1}
    W(0) &= 1, \nonumber \\
    X(0) &= (\epsilon_0 + p_0)e^{\nu_0 /2} \nonumber \\
    &\biggl\{ \biggl[ \frac{4\pi}{3} (\epsilon_0 + 3p_0) -\frac{\omega^2}{l} e^{-\nu_0} \biggr] W(0) +\dfrac{K(0)}{2} \biggr\}, \nonumber \\
    H_1(0) &= \dfrac{l K(0) + 8\pi (\epsilon_0 + p_0) W(0)}{n + 1},
\end{align}
\\
The condition $\Delta p = 0$ at $r=R$, the star's surface, requires the perturbation function $X$ to be equal to zero at the surface of the star. We assign some small arbitrary values to the functions $H_1$, $K$, and $W$ at the star's surface and integrate backward to reach the point at which the integration from the center of the star ends and join the two forward and backward solutions. 
 To find the quasinormal mode frequency for a given star, we solve the Zerilli equation given as \citep{Fackerell1971, Lindblom1983}
\begin{equation}
    d^2 Z / dr^{\star2} = \left(V_Z(r) - \omega^2 \right)Z(r^\star)
    \label{e:Zerrilli}
\end{equation}
where, $r^\star = r + 2 M log \left( r/2M -1 \right)$ and $V_Z(r)$ is the effective potential outside the star($r \geq R$).  The value of the Zerilli function $Z(r^\star)$, as given in Eq. 20 of \cite{Kunji2022}, depends only on $H_1$ and $K$ as the fluid perturbations $W$, $V$, and $X$ become non-existent outside the star. We calculate the value of $Z(r*)$ at the surface of the star from values of $H_1$ and $K$ at the surface. Outside the star Eq. \ref{e:Zerrilli} is numerically integrated  starting from the surface as done in \citep{Kunji2022, ZhaoLatt22} till a distance equivalent to $r = 25\omega^{-1}$. We match this value of $Z$ at $r = 25\omega^{-1}$ with that obtained from an asymptotic expansion of $Z$ valid far away from the surface of the NS. We impose the outgoing boundary condition which requires the amplitude of incoming gravitational wave far away from the surface of the star to be zero  \citep{Lindblom1983} and solve for $\omega$ as done in \citep{Kunji2022, ZhaoLatt22}.

\section{Results}
\label{sec:results}
Our primary aim is to comprehensively examine the $f-$mode oscillations, one of the non-radial oscillatory modes of NS over a broad range of realistic EOS. To achieve this, we employ the linearized and fully General Relativistic approach. For this work, we prepare a set of realistic nuclear EOS that only have nucleonic composition, obtained with the relativistic mean-field (RMF) approach. We utilize around 9,000 nucleonic EOS that were acquired through Bayesian inference with minimal restrictions imposed on them, as previously mentioned in Sec. \ref{RMF}. The range of nuclear saturation properties, including the binding energy $e_0$, nuclear incompressibility $K_0$, skewness parameter $Q_0$, symmetry energy $J_{\rm sym,0}$, its slope $L_{\rm sym,0}$ and its curvature $K_{\rm sym,0}$ at the saturation number density $n_0$, is presented in Table \ref{tab:nuc_params}. 
\begin{figure*}
    \includegraphics[width=\textwidth]{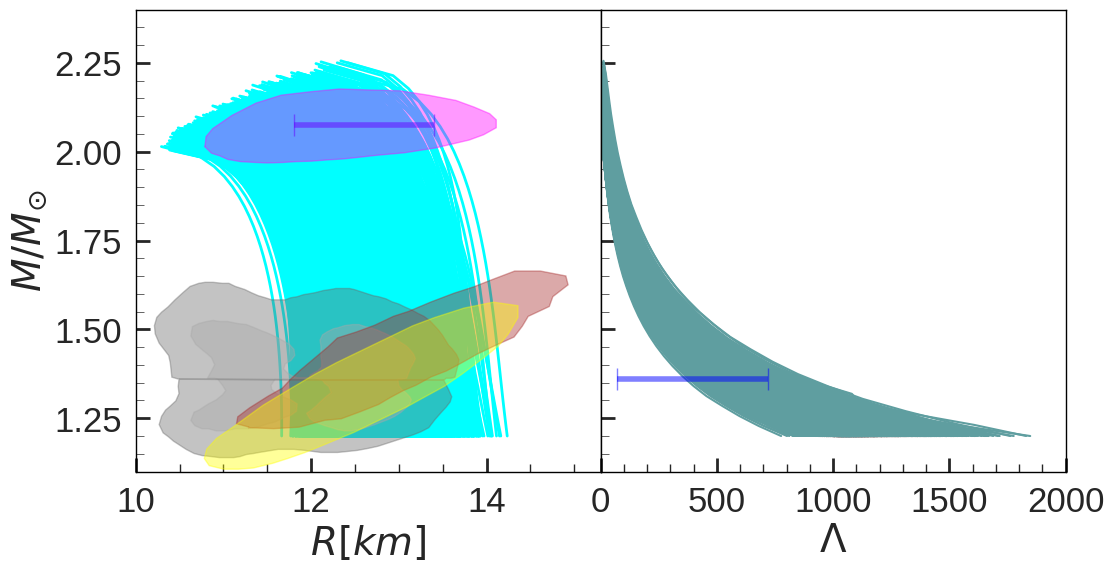}
    \caption{The NS mass-radius and mass-tidal deformability curves are plotted for the entire set of EOS used in this work. The gray zones in the left panel indicate  90\%(dark) and 50\%(light)  confidence intervals (CI) for the binary components of the GW170817 event \citep{Abbott1_2018}. The brown and yellow zones in the same panel represent the credible zone $1\sigma$ (68\%) of the 2-D posterior distribution in the mass-radius domain of millisecond pulsar PSR J0030+0451 in \citep{Riley2019} and \citep{Miller2019} respectively, while the magenta zone is for PSR J0740+6620 \citep{Riley2021, Miller2021} based on the NICER X-ray data. The blue bars represent the 90$\%$ CI radius of PSR J0740+6620 at 2.08$M_\odot$ (left panel) from \cite{Miller2021}
    and the tidal deformability from GW170817 event \citep{Abbott1_2018} at 1.36 $M_\odot$ (right panel).}
\label{fig:MR}
\end{figure*}
Our tabulated set of EOS is given a functional representation by spectral Decomposition Method (see Sec. \ref{subsec:spectral}). 
This is a reliable approach which is often used in the context of NS interior structure calculations. 
Using a minimal number of spectral coefficients, we achieved a fit of all our $\sim$ 9K EOS with a precision of less than 0.5\% using the Levenberg-Marquardt method for optimization of the residual error in Eq. \ref{e:residue_spectral} as done in \citep{Lindblom2010, Lindblom2018}.

\begin{table*}
\caption{Nuclear matter parameters calculated at saturation density for the 9K EOS used in this work, along with the median and 90\% confidence interval.}
\label{tab:nuc_params}
\setlength{\tabcolsep}{5.5pt}
\renewcommand{\arraystretch}{1.4}
\centering
\begin{tabular}{|ccccccccc|}
   \toprule
    Model & $e_0$ (MeV) & $n_0$ (fm$^{-3}$) & $K_0$ (MeV) & $Q_0$ (MeV) & $J_{sym,0}$ (MeV) & $L_{sym,0}$ (MeV) & $K_{sym,0}$ (MeV) & $m^\star$ \\
    \hline
    NL & $-16.10^{+0.33}_{-0.33}$ & $0.152^{+0.008}_{-0.005}$ & $255^{+42}_{-40}$ & $-451^{+154}_{-71}$ & $32.07^{+2.52}_{-2.77}$ & $32^{+13}_{-21}$ & $-96^{+96}_{-54}$ & $0.74^{+0.03}_{-0.10}$ \\
    \hline
\end{tabular}
\end{table*}

\begin{figure*}[t]
    \includegraphics[width= 0.45 \textwidth]{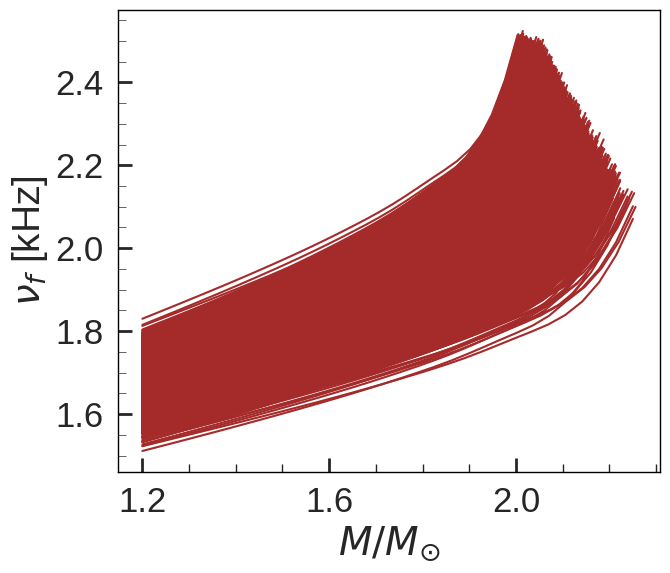}
    \includegraphics[width=0.415\linewidth]{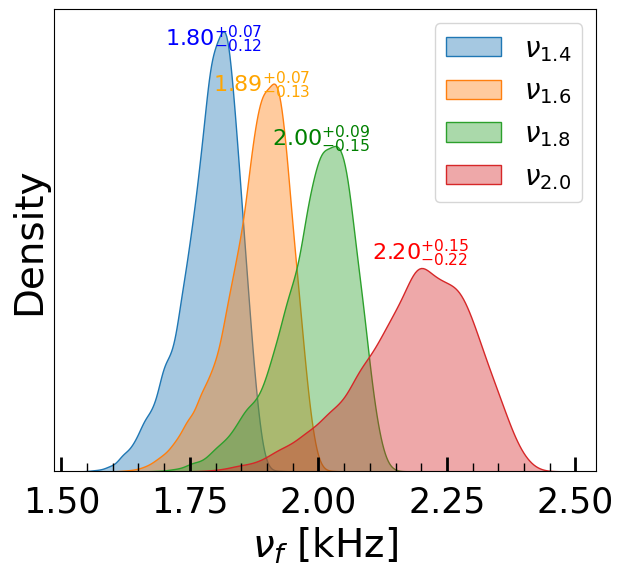}
    \caption{The frequency of the $f-$mode oscillations, calculated using linearized GR, is shown on the left for the EOS set utilized. On the right, the density distribution of the $f-$mode frequency for different NS masses in the range of 1.4 $M_\odot$ to 2.0 $M_\odot$ is plotted.}
\label{fig:f-modes}
\end{figure*}
In Fig. \ref{fig:EOS_pQCD}, we plot the entire EOS set employed in this work and compare with constraints derived from perturbative Quantum Chromodynamics (pQCD) at baryon densities of \(5 n_s\) and \(8 n_s\), with \(n_s=0.16 \, \text{fm}^{-3}\). The plot is divided into two panels, representing two different renormalization scales, (X = 1) and (X = 4), for the pQCD constraints. The left panel focuses on the renormalization scale (X = 1) of the pQCD constraints, with the green patch indicating the energy densities and pressure points for each EOS at the respective baryon densities that are in agreement with the pQCD constraints. 
All the EOS in our set meet the requirements set by pQCD results up to 8$n_s$ and a renormalization scale of 4. However, the central density for maximum-mass NS for all the EOS in our set is below 7$n_s$.

In Fig. \ref{fig:corner_spectral}, we present a corner plot showing the distribution of the spectral parameters for our set of EOS. This visual representation is essential to understand the statistical properties and correlations among the parameters. The diagonal plots contain representations of the marginalized 1D distribution of individual parameters, while the off-diagonal plots illustrate the 2D distribution for pairs of parameters, giving insight into their mutual relationships. 
The median value of each spectral parameter along with error is shown as a label on the corresponding diagonal plot and the dashed vertical lines mark the limits of the 68$\%$ confidence interval of the parameter.
The color gradient, which changes from dark to light blue in the 2D graphs, symbolizes the confidence intervals: 1$\sigma$, 2$\sigma$, and 3$\sigma$. The elliptical shapes observed in the 2D plots indicate strong correlations among the parameters. If the shape approaches a circle, it signifies a weak correlation between the given parameter pair.

An interesting observation from the figure is that most of the $\gamma_i$ parameters show strong correlations with their subsequent $\gamma_{i+1}$ parameters. This trend is expected; to ensure a smooth EOS, an increase in a particular $\gamma$ must be related to the previous one. This aspect of parameter correlation is essential for maintaining a coherent and smooth representation of the EOS, thus confirming the effectiveness of the spectral decomposition format in capturing the essential characteristics of the dense matter EOS.

In Fig. \ref{fig:MR}, we plot the mass-radius (MR) of the NS (left) and the mass-tidal deformability (right) of all EOS considered using the spectral representation of the EOS (see Sec. \ref{RMF}). The gray zones in the left panel represent the 90\% (dark) and 50\% (light) confidence intervals for the binary components of the GW170817 event \citep{Abbott2019properties}. The predictions of NS radius measurements from NICER X-ray data for the PSR J0030+0451 and PSR J0740+6620 pulsars are also included, with the $1\sigma$ (68\%) confidence zone of the 2-D posterior distribution in the MR domain of the millisecond pulsar PSR J0030+0451 shown in brown and yellow from references \citep{Riley2019} and \citep{Miller2019} respectively, and PSR J0740+6620 in magenta \citep{Riley2021, Miller2021}. Finally, the blue bar in the left panel represent the 90$\%$ CI radius of a 2.08$M_\odot$ NS as reported in \cite{Miller2021} combining the NICER data and X-ray observations. The blue bar in the right panel shows the bounds on the tidal deformability of a 1.36$M_\odot$ NS as given in \cite{Abbott1_2018}. All MR curves have a maximum mass greater than 2$M_\odot$, since this was imposed as a constraint during Bayesian sampling. Despite the dearth of constraints on radius and tidal deformability during sampling, the prediction of NS mass and radius are well within the observational bounds over the entire EOS set with the minimal requirements that were set. The obtained radius of a 1.4 $M_\odot$ NS in our set is 12.03-13.20 km, and the deformability of the tidal channel for a 1.36$M_\odot$ star is $431-659$ within 90\% confidence interval. The maximum mass of a NS ranges from 2 to 2.25$M_\odot$, which is consistent with the results reported in \cite{Rezzolla:2017aly}. It is worth noting that we applied perturbative Quantum Chromodynamics (pQCD) constraints to our EOS set, which limited the maximum mass of a NS to a certain range and eliminated some of the stiffer EOS.

We display the $f-$mode frequencies in Fig. \ref{fig:f-modes} (left) versus NS mass. It increases from 1.6 to 2.6 kHz for variation of NS mass from 1.2 to 2.25 $M_\odot$. It is worth noting that, the $f-$mode frequency within the Cowling approximation for NSs with a mass greater than $2M_\odot$ is in the range of 2.1 - 2.7 kHz and 2.3 - 2.65 kHz for the two types of EOS used in \citep{Kumar2023}, respectively. It is a widely known fact that the calculation of the $f-$mode frequencies using the Cowling approximation, overestimates them by up to 30 to 10 percent for NS with masses ranging from 1.0 to 2.5 M$_\odot$, when compared to the frequency obtained from the linearized General Relativistic formalism \cite{Benhar2004, Doneva2013, Pradhan2022}. In the right panel, we plot the distribution of $f-$ mode frequencies for different NS mass ranging from 1.4 to 2.0 $M_\odot$. Distribution for a given mass is accompanied by its median value and the 90\% confidence interval (CI) bounds. This variation in the $f-$mode frequencies across various NS masses is indicative of the current understanding of the EOS for NS. The measurements of $f-$mode frequencies (with precision beyond the current uncertainty of approximately 0.2 kHz) by the next generation detectors could potentially lead to a more refined understanding of the EOS domain for the NS interior.

\begin{figure}[b]
\centering
\includegraphics[width=0.45 \textwidth]{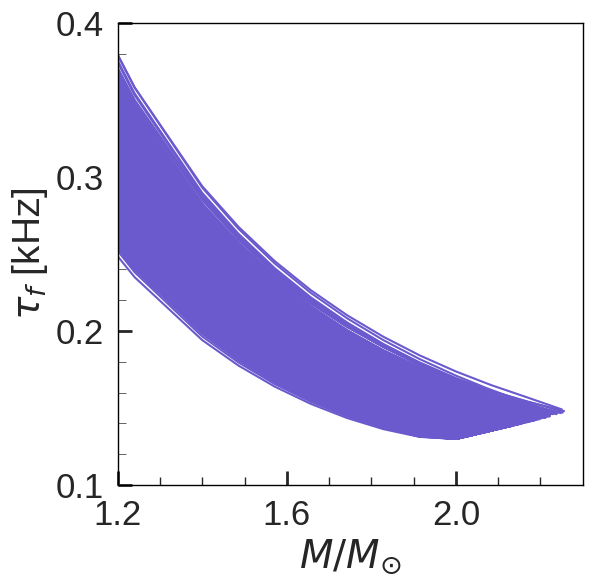}
\caption{Plot of the gravitational wave damping time, $\tau_f$ for the corresponding $f-$mode oscillation as a function of stellar mass for the 9K RMF EOS set used in this work.}
\label{fig:tau}
\end{figure}

Next we calculate  the gravitational wave damping times corresponding to the $f-$mode frequencies for our EOS set. The inverse of the imaginary part of $\omega$ in the metric perturbations gives us the damping time as discussed in Sec.\ref{omega}. In Fig. \ref{fig:tau} we plot the variation of damping times with stellar mass. We obtain a monotonically decreasing trend for the damping times with increase in NS mass till the maximum mass for all the EOS. 
The damping time of a 1.4 $M_\odot$(2.0 $M_\odot$) NS lies between 0.20 - 0.25 s (0.13 - 0.15 s). 
The range of damping times in this work, for the two NS masses mentioned is in good agreement with the range of damping times reported in \cite{Pradhan2022} for their set of RMF EOS.

\begin{figure*}
\includegraphics[width=\textwidth]{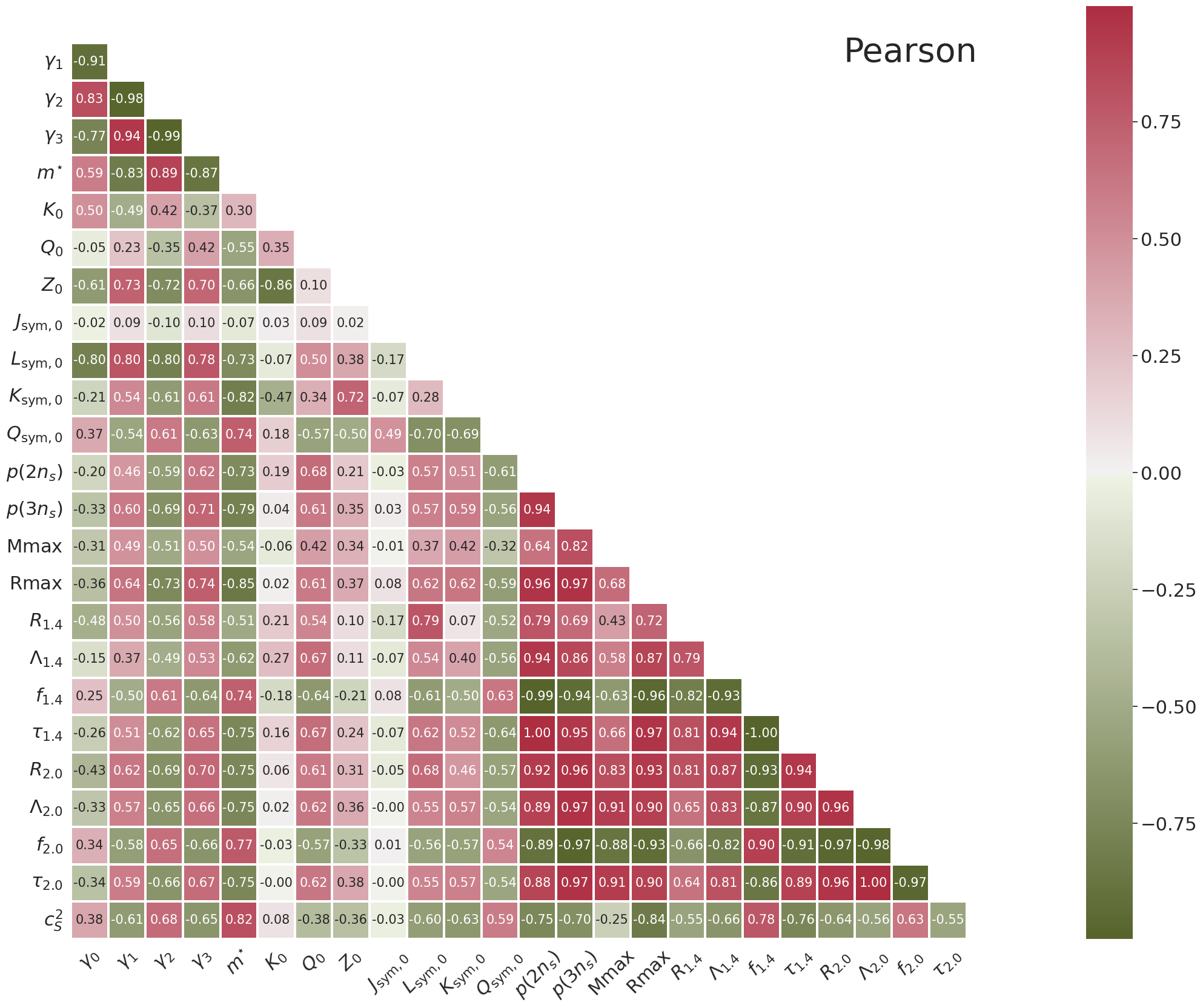}
\caption{Heatmap displaying the Pearson correlation coefficients among the spectral parameters ($\gamma_0$, $\gamma_1$, $\gamma_2$, $\gamma_3$), nuclear matter parameters ($m^\star$, $K_0$, $Q_0$, etc.), pressure at multiples of saturation number density $n_s$ = 0.16 fm$^{-3}$, and stellar observables like mass, radius, $f-$mode frequency, etc. Subscripts with the corresponding NS property stand for the NS mass associated with that property.}
\label{fig:corr_heatmap}
\end{figure*}

\begin{figure*}[t]
    \includegraphics[width=\textwidth]{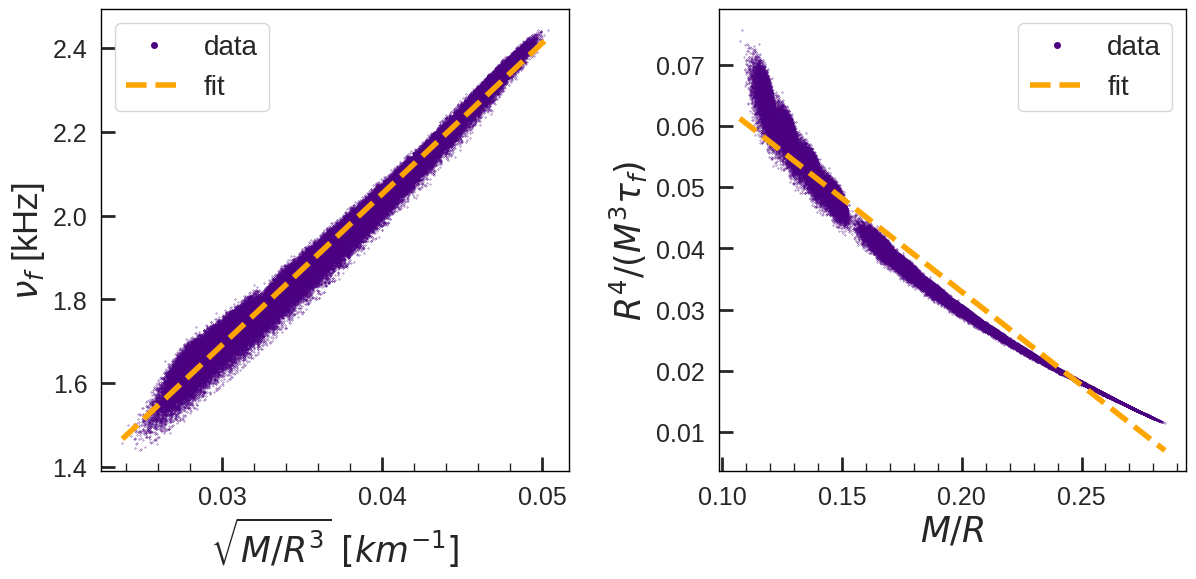}
    \caption{Empirical relation between the $f-$mode frequency with the average stellar density (left panel) and that between the corresponding dimensionless gravitational wave damping time and the stellar compactness for the set of 9K RMF EOS used in this work. Stellar mass $M$, radius $R$, and damping time $\tau_f$ are in km.}
    \label{fig:nu_and_tau_UR}
\end{figure*}
In the following, we explore the linear dependence between various spectral parameters, nuclear saturation properties, the pressure of NS matter at two and three times the fixed baryon density $n_s$, and other NS properties. We also examine how the frequency of the $f-$mode oscillation for different masses relates to the individual components of our nuclear EOS set. We calculate the Pearson correlation coefficients between these quantities using the entire $\sim9$K EOS set. The heatmap's color spectrum, ranging from green to red, indicates the strength of the correlation, from negative to positive, respectively. A deeper red(green) hue indicates a stronger positive(negative) correlation, with a magnitude of coefficients above 0.8 indicating a particularly strong linear relationship. The correlation between the spectral parameters $\gamma_1$, $\gamma_2$ and $\gamma_3$ is exceptionally high, clearly indicating a linear relationship between them. This is also true for the pair $m^\star$ and the $\gamma_i$, suggesting that the effective mass of the nucleon has a strong influence on these spectral parameters. Another finding is the strong correlation, greater than 0.9, between $P (2n_s)$ and $P (3n_s)$, which are the pressures $\beta$ at twice and three times of $n_s$ and various properties of NS such as radius, deformability of the tidal, frequency of the mode $f$ and damping time for NS masses of 1.4 $M_\odot$ and 2.0 $M_\odot$. This suggests a considerable degree of linear dependence between the properties of NS of different masses and the overall EOS, rather than individual nuclear saturation properties.
It is worth emphasizing the strong correlation previously reported in Ref. \cite{Pradhan2022} between the fundamental $f-$mode frequency of different NS masses and the effective nucleon mass at saturation $m^\star$, appears to be less pronounced in our study using the chosen set of EOS. This difference can be mainly attributed to certain choices regarding the set of EOS. The set employed in our research was generated through Bayesian inference subject to constraints. Specifically, constraints from chiral effective field theory ($\chi$EFT) for our low density pure neutron matter and perturbative quantum chromodynamics (pQCD) at 8 times $n_s$ were included. These constraints were not considered by the other authors. An additional notable distinction is the inclusion of an interaction $\omega^4$ in the Lagrangian that we studied. It is important to note that this term $\omega^4$ tends to soften the EOS, which is a characteristic favored by pQCD analysis \citep{Malik2023}. The heatmap suggests that the maximum radius of a NS is closely related to its $R_{1.4}$, $R_{2.0}$, and $f_{2.0}$ values. Furthermore, the $f-$mode frequencies of NS with 1.4 $M_\odot$ and 2.0 $M_\odot$ are strongly correlated with their tidal deformabilities $\Lambda_{1.4}$, and $\Lambda_{2.0}$, respectively, providing insight into how the internal structure of NS responds to tidal forces.

Asteroseismology, the study of stellar oscillations, allows us to infer global NS properties such as mass, radius, and average stellar density from $f-$mode frequencies and the gravitational wave damping time through certain empirical relations. We describe the results for some selected relations in the following paragraphs.

In Fig. \ref{fig:nu_and_tau_UR} we plot the data with the straight line fit using:
\begin{equation}
    \nu_f = a \sqrt{M/R^3} + b
    \label{e:average_density}
\end{equation}
This empirical relation is well-known since the publication of \cite{andersson1998}. We find the slope $a$ to be $35.94 \pm 0.0113$ kHz.km and the intercept $b$ to be $0.6260 \pm 0.0004$ kHz. We can compare these values with the values reported in \citep{Pradhan2022} (See Table \ref{tab:average_density}). They calculated the $f-$mode frequencies with a different set of RMF EOS and found the slope equal to $36.20$ kHz.km and the intercept equal to $0.535$ kHz. \cite{Kumar2023} report the same coefficients with values: $a=22.27,~26.76$ and $b=1.520,~1.348$ corresponding to determination of $f-$modes with Cowling approximation for a set of hybrid and a set of purely nucleonic EOS with density dependent couplings. The values for $a$ and $b$ thus appears to be dependent on the set of EOS considered and makes the empirical Eq. \ref{e:average_density} somewhat EOS dependent. The relative percentage error for the aforementioned fit in this work is $2.84\%$.
\begin{table*}
    \centering
    \begin{tabular}{| m{2.5cm} m{1.5cm} m{3cm} m{1.2cm} |m{3.5cm} m{3.5cm}|}
        \hline
\multicolumn{4}{|c|}{[FullGR]}&[FullGR] & [Cowling] \\
        Empirical & Fitting  & Coefficients & Relative &Coefficients&Coefficients\\
         relation & function & of current work & error & of \protect\citep{Pradhan2022}&of \protect\citep{Kumar2023} \\
        \hline
        $f~-~\sqrt{M/R^3}$ & $ax + b$ & $a = 35.94 \pm 0.0113$ & $2.84\%$ & $a = 36.20$ & $a = 22.27,~26.76$ \\
         &  & $b = 0.6260 \pm 0.0004$ & & $b = 0.535$& $b = 1.520,~1.348$\\
         \hline
    \end{tabular}
    \caption{Coefficients of the fitting equation for the empirical relations Eq.\ref{e:average_density}}
    \label{tab:average_density}
\end{table*}

We display another empirical relation:
\begin{equation}
    \dfrac{R^4}{M^3 \tau_f} = a \frac{M}{R} + b,
    \label{e:tau_f}
\end{equation}
in the right panel of Fig. \ref{fig:nu_and_tau_UR}. $\tau_f$ is the gravitational wave damping time, equal to the inverse of the imaginary part of $\omega$. \cite{andersson1998} mentioned the existence of this scaling between a dimensionless damping time and the compactness for the first time. Both the coefficients $a$ and $b$ are dimensionless and we find the following values (See Table \ref{tab:tau_f}):  $a = -0.3051 \pm 0.0001$, $b = 0.0939 \pm 3.0 \times 10^{-5}$ with a relative error of $0.29\%$. \citep{Pradhan2022} reported the values: $a = -0.245$ and $b = 0.080$. These empirical relations have poor validity beyond $M/R > 0.25$ \citep{tsui2005universality}.

\begin{table*}
    \centering
    \begin{tabular}{| m{3.0cm} m{1.5cm} m{3.5cm} m{1.2cm} |m{3.5cm}|}
        \hline
\multicolumn{4}{|c|}{[FullGR]}&[FullGR]  \\
        Empirical & Fitting  & Coefficients & Relative & Coefficients \\
         relation & function & of current work & error & of \citep{Pradhan2022}\\
        \hline
        \vspace{0.1cm}
        $\dfrac{R^4}{M^3 \tau_f}~-~M/R$ & $ax + b$ & $a = -0.3051 \pm 0.0001$ & $0.29\%$ & $a = -0.245$ \\
         &  & $b = 0.0939 \pm 3.0 \times 10^{-5}$ & & $b = 0.080$ \\
         \hline
    \end{tabular}
    \caption{Coefficients of the fitting equation for the empirical relations Eq.\ref{e:tau_f}}
    \label{tab:tau_f}
\end{table*}

\begin{figure}
    \includegraphics[width=0.5 \textwidth]{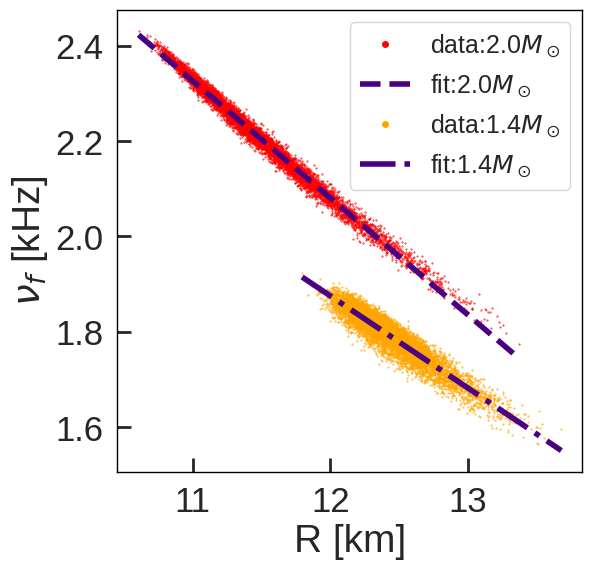}
    \caption{The empirical relation between the radii and the corresponding $f-$mode frequencies for the set of 9K RMF EOS used in this work. The orange(red) dots denote $f-$mode frequencies for NS with 1.4(2.0) $M_\odot$. The dashed-dot(dashed) line represents the linear fit based on Eq. \ref{e:fmode_radius_UR} for data for 1.4(2.0) $M_\odot$.}
    \label{fig:fmodes_radius_UR}
\end{figure}

\cite{Kumar2023} have reported a new scaling between the $f-$mode frequency and the radius for a NS of given stellar mass:
\begin{equation}
    f_x = aR_x + b,
    \label{e:fmode_radius_UR}
\end{equation}
where $x$ is the mass of the NS in $M_\odot$ units. We plot the relation with our results of $f-$mode frequencies for 1.4 $M_\odot$ and 2.0 $M_\odot$ in Fig. \ref{fig:fmodes_radius_UR}.

We find the values (See Table \ref{tab:fmode_radius_UR}): $a = -0.1933 \pm 0.0008$ kHz/km, $b = 4.195 \pm 0.010$ kHz for 1.4 $M_\odot$ and $a = -0.2455 \pm 0.0003$ kHz/km, $b = 5.027 \pm 0.003$ kHz for 2.0 $M_\odot$. The maximum relative errors are $1.86\%$ and $1.18\%$ respectively. The slope reported in \cite{Kumar2023} for masses over the range 1.6 $M_\odot$ - 2.4 $M_\odot$ is around $-0.22$ kHz/km and the intercept is around $5.1$ khz with $1.5\%$ relative error. A general relativistic treatment of the stellar oscillations in this work presents us with two different slopes for the two cases. This difference is due to the overestimation of $f-$mode frequencies that occurs when one employs Cowling approximation, as done in \cite{Kumar2023}. From a specific observed value of $f-$mode frequency for 1.4 $M_\odot$, we can infer a lesser value of radius from the fit in this work as compared to the fit in \cite{Kumar2023}. 

\begin{table*}
    \centering
    \begin{tabular}{| m{3.0cm} m{1.5cm} m{3.5cm} m{1.2cm} |m{3.5cm}|}
        \hline
\multicolumn{4}{|c|}{[FullGR]} & [Cowling] \\
        Empirical & Fitting  & Coefficients & Relative & Coefficients \\
         relation & function & of current work & error & of \citep{Kumar2023}\\
        \hline
        $f_{1.4}~-~R_{1.4}$ & $ax + b$ & $a = -0.1933 \pm 0.0008$ & $1.86\%$ &  \\
         &  & $b = 4.1949 \pm 0.0095$ & & $a = -0.22$ \\
         $f_{2.0}~-~R_{2.0}$ & $ax + b$ & $a = -0.2455 \pm 0.0003$ & $1.18\%$ & $ b = 5.1$\\
         &  & $b = 5.0268 \pm 0.0034$ & & \\
         \hline
    \end{tabular}
    \caption{Coefficients of the fitting equation for the empirical relations Eq.\ref{e:fmode_radius_UR}}
    \label{tab:fmode_radius_UR}
\end{table*}

\begin{figure*}
    \centering
    \includegraphics[width=0.99\linewidth]{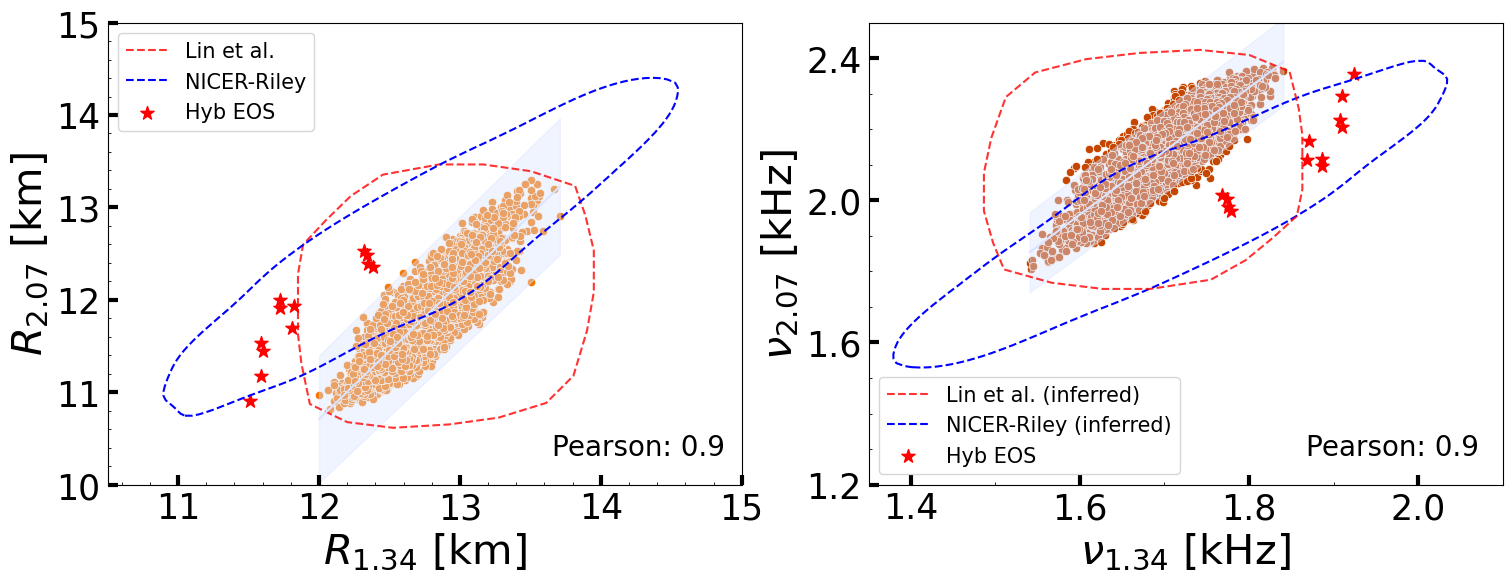}
    \caption{The relationship between the radii (left panel) and the $f-$mode frequencies (right panel) of NS with masses of 1.34 $M_\odot$ and 2.07 $M_\odot$ for the EOS set used in this work. The dashed blue line on the left shows the NICER constraint, which was determined by taking the 1$\sigma$ posteriors of the radius values from NICER measurements of PSR J0740+6620 and PSR 0030+0451 and marginalizing them over the NS mass while taking into account the correlation between them, based on the data in \cite{Riley2019, Riley2021}.
    The area inside the red dashed boundary is the most probable region of the joint posterior distribution of the NICER radii measurements as in \cite{Lin:2023cbo}. On the right, we inferred $f-$mode frequencies from the data in the radius domain using the universal relation in Eq. \ref{e:fmode_radius_UR} (see text for details). The red stars in the left panel(right panel) mark the calculated values corresponding to radii($f-$mode  frequencies) for a set of hybrid EOS \citep{Baym2018, Baym2019, Kojo2022}, which support hadron-quark phase transition.}
    \label{fig:rf_cor_Riley}
\end{figure*}

Motivated by the study in \cite{Lin:2023cbo}, we also look into the correlation between the radii as well as $f-$mode frequencies of two extreme mass NS, utilizing our nucleonic EOS set.
In Fig. \ref{fig:rf_cor_Riley}, we show a scatter plot of the radii ($f-$mode frequencies) of NS with masses of 1.34 $M_\odot$ and 2.07 $M_\odot$ on the left (right). On both the panels, a strong correlation is noticed, with a Pearson's correlation coefficient of approximately 0.9, for our EOS set. In the left panel, the red dashed line indicates the most probable region of the joint posterior distribution of the NICER radii measurements of PSR J0740+6620 and PSR 0030+0451 calculated in \cite{Lin:2023cbo}. The authors have claimed a 48\% probability (with a 5\% false alarm rate) of identifying a strong and sharp phase transition by applying their method on NICER observations. We note how our entire nucleonic set overlaps with the most probable region within the red dashed line. In the same panel, the area marked with blue dashed line represents the constraints after we impose a strong correlation of 0.9 among the radius data in the NICER observations of the pulsars PSR J0740+6620 and PSR 0030+0451. NICER's $1\sigma$ posterior distributions are used to obtain this constraint. This method is based on data reported in \cite{Riley2019, Riley2021}. To be more specific, we produce 2000 random samples of $R_{1.34}$ and $R_{2.07}$ within the $1\sigma$ range of the NICER's observed values marginalized over NS masses and implemented a correlation of approximately 0.9 between the two radii. The plot shows a noticeable difference in the slope of the data points corresponding to nucleonic EOS and those derived from NICER observations.

\begin{figure*}
    \centering
    \includegraphics[width=0.99\linewidth]{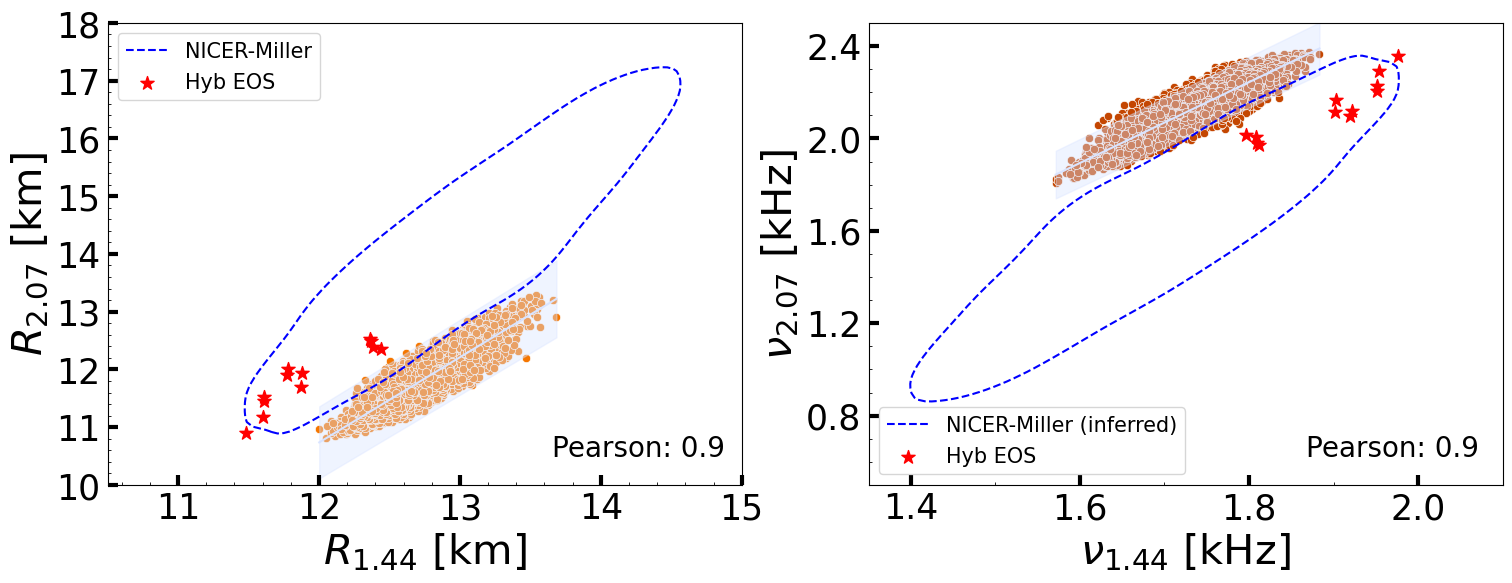}
    \caption{The relationship between the radii (left panel) and the $f-$mode frequencies (right panel) of NS with masses of 1.44 $M_\odot$ and 2.07 $M_\odot$ for the EOS set used in this work. The dashed blue line on the left shows the NICER constraint, which was determined by taking the 1$\sigma$ posteriors of the radius values from NICER measurements of PSR J0740+6620 and PSR 0030+0451 and marginalizing them over the NS mass while taking into account the correlation between them, based on the data in \cite{Miller2019, Miller2021}.
    On the right, we inferred $f-$mode frequencies from the data in the radius domain using the universal relation in Eq. \ref{e:fmode_radius_UR} (see text for details). The red stars in the left panel(right panel) mark the calculated values corresponding to radii($f-$mode  frequencies) for a set of hybrid EOS \citep{Baym2018, Baym2019, Kojo2022}, which support hadron-quark phase transition.}
    \label{fig:rf_cor_Miller}
\end{figure*}

In the right panel, we translate the NICER data from \cite{Riley2019, Riley2021} as well as that from \cite{Lin:2023cbo}, in the radius domain using Eq. \ref{e:fmode_radius_UR} to obtain the corresponding constraints in the $f-$mode frequency domain. Note that the universal relation presented in Table 4 applies to NS masses of 1.4 $M_\odot$ and 2.0 $M_\odot$. For Fig. \ref{fig:rf_cor_Riley} we calculate the relations for NS masses of 1.34 $M_\odot$ and 2.07 $M_\odot$. The slope and the intercept values are very close to those presented in Table 4: -0.1794 (-0.2360) kHz/km and 3.9891 (4.9283) kHz for NS masses of 1.34 (2.07) $M_\odot$, respectively. The nucleonic EOS set exhibits a similar overlap between the radius and frequency domains with the NICER-inferred constraint from the data in \cite{Riley2019, Riley2021}. However, the partial overlap between the nucleonic EOS set and this NICER bound suggests that only nucleonic degrees of freedom for larger NS masses may not be sufficient enough to explain the observations.\\
\\
In Fig. \ref{fig:rf_cor_Miller}, we present a similar analysis as in Fig. \ref{fig:rf_cor_Riley} for NS masses 1.44 $M_\odot$ and 2.07 $M_\odot$ but determine the NICER constraints from data published by \cite{Miller2019, Miller2021}. The slope and the intercept values for a 1.44 $M_\odot$ are -0.1869 kHz/km and 4.1215 kHz respectively. We observe a clear distinction between the NICER bound and the nucleonic EOS set in both the radius and frequency domains. This finding further strengthens the conclusion drawn from the previous figure.\\
\\
To investigate further, we utilize 12 EOS models with NS maximum mass above 2.0 $M_\odot$ simulating hadron-quark phase transition inside the NS core from the CompOSE database, such as VQCD\citep{Jokela2021}, QHC21-AT\citep{Kojo2022}, QHC19-C\citep{Baym2019}, QHC21-BT, QHC21-D, QHC19-D, QHC21-CT, QHC21-B, QHC21-C, QHC18\citep{Baym2018}, QHC19-B, QHC21-DT. The red stars in both Fig. \ref{fig:rf_cor_Riley} and Fig. \ref{fig:rf_cor_Miller} with the label 'Hyb-EOS' in both the panels are calculated values for these EOS in the corresponding domains. It is clear that all the red stars fit comfortably within our NICER-derived constraints, as determined by both - Riley et al. and Miller et al., over the radius and the $f-$mode frequency domain. The possibility of a hadron-quark phase transition or other exotic degrees of freedom in the NS EOS appears to be promising.
Please be aware that the blue dashed contours derived from the NICER data for the pulsars: J0030+0451 and PSR J0740+6620, whether in the radius domain or in the $f-$mode frequency domain, is calculated using a correlation of 0.9 between the radii of NS with these two masses, and is entirely dependent on the nucleonic degrees of freedom. Examining the differences in these correlations with other exotic degrees of freedom is a subject for future research and is not within the purview of this study.\\
\\
The universal relation in Eq. \ref{e:fmode_radius_UR} acts as a tool to compare data from two different channels: X-ray channel and gravitational wave channel.  Observation of thermal emission from hotspots on pulsars provide an effective way of measuring the NS radius. Parameter estimation of parametrized models of NS EOS using such radius measurements \citep{Raaijmakers2021} allows us to narrow down viable EOS by putting constraints on the relevant parameters. Radius measurements from X-ray data are essentially based on X-ray pulse profile modelling \citep{Bogdanov2021}. This involves making assumptions about the stellar atmosphere and the equation of state of matter inside the star itself. In short, such measurements are model dependent. $f-$ mode observations offer a more direct path to probe the NS EOS by measuring the bulk properties of the star through its response to gravity. Thus, X-ray observations, while valuable, require additional assumptions about the star's surface and internal structure, making them less direct for studying the deep core where the EOS is truly uncertain as yet. These observations work best together \citep{Yunes2022}, with X-rays providing complementary data points like mass and radius to refine the interpretation of the NS EOS extracted from gravitational waves \citep{Psaltis2014}.\\
\vspace{1cm}

\section{Conclusion}
\label{sec:conclusion}
We have carefully calculated the quasinormal mode frequency, specifically focusing on the $f-$mode oscillations of NS. This analysis was conducted using a linearized approach to full General Relativity, which allowed for a detailed examination of these oscillations. In our study, we utilized a wide range of nuclear EOS (9K), derived through Bayesian inference. These EOS were subjected to few constraints. Firstly, they conform to the minimal constraints on nuclear saturation properties. Secondly, they incorporate constraints from chiral effective field theory ($\chi$EFT) relevant to low-density pure neutron matter. Additionally, constraints derived from perturbative Quantum Chromodynamics (pQCD) at densities pertinent to the core of NS were also applied. Finally, our selection of EOS ensured that the NS maximum mass was above 2$M_\odot$, aligning with current astrophysical observations and theoretical predictions.

We then determine the spectral representation of our EOS set, which is one of the most widely accepted and optimal way to adapt any realistic EOS table into a functional form. This helps to reduce the computation time when calculating NS properties and is convenient for other astrophysical calculations. The mass-radius and mass-tidal deformability relationships of NS were calculated, and a detailed examination of the correlation among spectral parameters, various nuclear saturation properties and NS properties was conducted. Empirical relations for $f-$mode frequencies and gravitational wave damping times were also explored, providing valuable insights into the composition of these compact objects.

One of the key findings is that the $f-$mode frequencies do not show a strong relationship with any single nuclear saturation property of the EOS when the 9K EOS is used. In fact, we find that some of the strong correlations reported in recent publications have become weak for our EOS set. This observation suggests that a more nuanced approach, possibly involving multiparameter correlation analysis using machine learning techniques, might be necessary to unravel these complex relationships.\\
\\
Interestingly, for NS with masses: $\sim$1.4 $M_\odot$ and $\sim$2.0 $M_\odot$, we notice a strong correlation between their radii as well as $f-$mode frequencies.  Using this correlation alongside NICER observations of PSR J0740+6620 and PSR 0030+0451, we find partial and minimal overlap for observational data from Riley et al. and Miller et al. respectively with our nucleonic EOS dataset. This motivates us to further investigate these results for the presence of non-nucleonic degrees of freedom in the EOS. Hence, we employ a selection of 12 hybrid EOS from the CompOSE database to calculate the radius and the $f-$mode frequency data which appear to align more closely with the above-mentioned NICER constraints particularly in regions where the nucleonic EOS shows no overlap. Note, that the NICER constraints were derived on the basis of a correlation of 0.9 among the radii of 1.4 and 2.07 M$_\odot$ NS calculated with our nucleonic set. Thus our findings emphasise the need to systematically check for a signature of a hadron-quark phase transition in the NS interior.\\
\\
A recent study by \cite{Lin:2023cbo} also demonstrates the correlation between the radii of NS for the observed masses of the same pulsars, as detected by NICER. Their analysis suggests the existence of a sharp and strong phase transition with 48\% probability.
Although our methodology differs from that of \cite{Lin:2023cbo}, our conclusions are similarly aligned. Our study provides a clearer distinction between nucleonic and non-nucleonic EOS across two domains over NICER data from two complimentary analyses. By translating NICER radius measurements into the $f-$mode frequency domain using a universal relationship, we highlight an alternative avenue through gravitational wave astronomy to compare various EOS models and check for the presence of exotic phases through the gravitational wave channel.\\
\\
For the first time, our analysis shows that the NICER constraints on radius prefer hybrid EOS over purely nucleonic ones, which also holds true in the $f-$mode frequency domain. It is also expected that NS with higher mass for example 2.0 $M_\odot$, will have a different composition rather than 1.4 $M_\odot$ NS. The conditions inside a higher mass NS with higher central density than that inside a 1.4 $M_\odot$ NS are more favourable for a deconfined quark phase.  We expect future precise measurements of the $f-$mode frequency, particularly those for NS with masses near the two extremes of viable NS mass values, will provide decisive evidence regarding the internal composition of NS. Such measurements will be able to differentiate between purely nucleonic matter and other exotic phases, shedding light on the state of matter at extreme densities.
We would do a thorough and organized study along similar lines taking into account a wide range of hybrid EOS generated using multiple phenomenological models and report our results in a future communication.

\section*{ACKNOWLEDGMENTS} 
T.M. would like to acknowledge the support from national funds from FCT (Fundaço para a Ciência e a Tecnologia, IEP, Portugal) under Projects No. UIDP/-04564/-2020, No. UIDB/-04564/-2020 and 2022.06460.PTDC. T.M. is also grateful for the support of EURO-LABS "EUROpean Laboratories for Accelerator Based Science", which was funded by the Horizon Europe research and innovation program with grant agreement No. 101057511. D. G. R., Sw.B., and Sa.B. would like to acknowledge the financial support by DST-SERB, Govt. of India through the Core Research Grant (CRG/2020/003899) for the project entitled: `Investigating the Equation of State of Neutron Stars through Gravitational Wave Emission'. Finally, D. G. R. would like to thank colleagues - A. Venneti and K. Nobleson for numerous suggestions regarding the work done in this publication.

\bibliographystyle{aasjournal}
\bibliography{refs}
\end{document}